\documentclass[aps,prd,showpacs,nofootinbib,floats,floatfix,reprint,preprintnumbers,groupedaddress,twocolumn]{revtex4}

\usepackage{bm}
\usepackage{latexsym}
\usepackage{dcolumn}
\usepackage{amsmath,amsfonts,amssymb}
\usepackage{graphicx,epsfig}
\usepackage{amsmath}
\usepackage{fancyhdr}
\usepackage{hyperref}
\usepackage{graphicx,epstopdf}
\hypersetup{
	colorlinks   = true, %Colours links instead of ugly boxes
	urlcolor     = blue, %Colour for external hyperlinks
	linkcolor    = blue, %Colour of internal links
	citecolor   = red %Colour of citations
}

\begin{document}
\title{Study of relativistic accretion flow in Kerr-Taub-NUT spacetime}
\author{Indu K. Dihingia$^{1,2}$}\email{idihingia@iiti.ac.in }
\author{Debaprasad Maity$^{1}$}\email{debu@iitg.ac.in }
\author{Sayan Chakrabarti$^{1}$}\email{sayan.chakrabarti@iitg.ac.in}
\author{Santabrata Das$^{1}$}\email{sbdas@iitg.ac.in }

\affiliation{$^1$Department of Physics, Indian Institute of Technology Guwahati, Guwahati 781039, Assam, India\\
$^2$Discipline of Astronomy, Astrophysics and Space Engineering, Indian Institute of Technology Indore, Indore 453552, India
}

\date{\today}

\begin{abstract}
	We study the properties of the relativistic, steady, axisymmetric, low angular momentum, inviscid, advective, geometrically thin accretion flow in a Kerr-Taub-NUT (KTN) spacetime which is characterized by the Kerr parameter ($a_{\rm k}$) and NUT parameter ($n$). Depending on $a_{\rm k}$ and $n$ values, KTN spacetime represents either a black or a naked singularity. We solve the governing equations that describe the relativistic accretion flow in KTN spacetime and obtain all possible global transonic accretion solutions around KTN black hole in terms of the energy $({\cal E})$ and angular momentum $(\lambda)$ of the flow. We identify the region of the parameter space in $\lambda-{\cal E}$ plane that admits the flow to possess multiple critical points for KTN black hole. We examine the modification of the parameter space due to $a_{\rm k}$ and $n$ and find that the role of $a_{\rm k}$ and $n$ in determining the parameter space is opposite to each other. This clearly indicates that the NUT parameter $n$ effectively mitigate the effect of black hole rotation in deciding the accretion flow structure. Further, we calculate the disc luminosity ($L$) corresponding to the accretion solutions around the KTN black hole and for a given set of $a_{\rm k}$ and $n$, we obtain the maximum luminosity $(L_{\rm max})$ by freely varying $\lambda$ and ${\cal E}$. We observe that $L_{\rm max}$ decreases with the increase of $n$ irrespective of $a_{\rm k}$. In addition, we also investigate all possible flow topologies around the naked singularity and find that there exists a region around the naked singularity which remains inaccessible to the flow. We study the critical point properties for naked singularities and find that the flow possesses maximum of four critical points. Finally, we obtain the parameter space for multiple critical points for naked singularity and find that parameter space is shrunk and shifted to lower $\lambda$ and higher ${\cal E}$ side as $a_{\rm k}$ is increased which ultimately disappears.
	
\end{abstract}

\pacs{-----------------}
\maketitle

\section{Introduction}

The accretion process around the compact stars remains the subject of intense interest for last several decades in the astrophysical community. Understanding the electromagnetic properties of a large class of astrophysical observations, particularly for the sources like quasars, active galactic nuclei and black hole X-ray binaries, the accretion of matter has been proved to be the potentially possible physical mechanism till date \cite{Shakura:1972te,1985aagq.conf..411B,Blaes:2007zm, netzer_2013,  Abramowicz:2011xu}. Generically, black holes are considered to be the central object which are essentially a very special class of solutions of the well known Einstein's equation. One of the defining properties of black holes is the existence of the horizons which is a surface that encompasses the curvature singularity in a black hole spacetime and the horizon behaves like a one way membrane through which anything can enter but nothing can come out. This interesting property helps one to invoke unique boundary condition for the accretion flow dynamics near the horizon. The underlying framework of studying the accretion of matter is based upon the principles of relativistic hydrodynamics in gravitational background. Once the flow properties, such as velocity, temperature, density etc. are understood, the relevant radiative processes can be computed and compared with the observation. Therefore, in principle, one can put constraints on the underlying theoretical parameters, such as the mass accretion rate as well as the mass and spin of the black hole.

Vast amount of literature exists on the topic of accretion flow which are based on different physical conditions in the hydrodynamic regime (see \cite{Abramowicz:2011xu,Font_2000} and the references therein). However, limited works involving hydrodynamical aspects of accretion flow have been performed in the realm of modified gravitational backgrounds as well as around the exotic compact objects in general relativity. 
%little attention is paid so far in studying such flow dynamics in the realm of modified gravitational backgrounds. 
For instance, the accretion flows around brane-world black holes \cite{PhysRevD.78.084015, HeydariFard:2010zza}, slowly rotating black holes in dynamical Chern-Simons modified gravity \cite{Harko_2010}, black holes in Ho\v{r}ava gravity \cite{Harko:2010ua, PhysRevD.80.044021}, boson stars \cite{Torres_2002,PhysRevD.73.021501}, wormholes \cite{PhysRevD.79.064001}, gravastars \cite{Harko:2009gc}, and quark stars \cite{Kov_cs_2009}, etc., were studied considering particle dynamics.
%For instance, the properties of the accretion flow were investigated for exotic central objects, such as boson stars \cite{Torres_2002,PhysRevD.73.021501}, wormholes \cite{PhysRevD.79.064001}, gravastars \cite{Harko:2009gc}, brane-world black holes \cite{PhysRevD.78.084015, HeydariFard:2010zza}, slowly rotating black holes in dynamical Chern-Simons modified gravity \cite{Harko_2010}, black holes in Ho\v{r}ava gravity \cite{Harko:2010ua, PhysRevD.80.044021} and quark stars \cite{Kov_cs_2009}, \textcolor{red}{etc}. 
In addition, accretion disc properties have been studied around naked singularities \cite{1979ApJ...227..596R,Joshi_2013,Kov_cs_2010} as well. 

It may be noted that the astrophysical observations are generally explained considering various background gravitational systems which may comprise of black hole or any other exotic compact objects. Therefore, given the advent of high precession observations these days, it would be viable to probe the nature of gravitational background through the study of the accretion flow dynamics, and this is the main motivation of our present study. 
%The goal can be achieved by retaining ourselves within the realm of Einstein's general theory of relativity or going beyond it. 
Towards this, for the first time to the best of our knowledge, we study the properties of the general relativistic accretion flow around the general class of gravitational backgrounds which are the solutions of vacuum Einstein's equation. Here, we emphasize that depending upon the choice of parameters of the theory, those classes of gravitational backgrounds do represent either black holes or more exotic spacetimes with naked singularity. 
 
Historically, for the first time, \citet{10.2307/1969567} reported a gravitational background which is presently known as the Taub-NUT spacetime. The initial motivation to construct such a spacetime was based on the assumption of the existence of a four-dimensional group of isometries such that the spacetime can be interpreted as a possible vacuum homogeneous cosmological model. Thereafter, the solution was rediscovered by Newman, Unti and Tamburino (NUT) \cite{doi:10.1063/1.1704018} as a simple generalization of the Schwarzschild spacetime. To include rotation, the Kerr-Taub-NUT (KTN) spacetime was formulated that generalizes the well known Kerr metric by introducing a new parameter called the NUT charge.
%The Kerr-Taub-NUT (KTN) spacetime, on the other hand, is a class of gravitational background which generalizes the well known Kerr metric by introducing a new parameter called the NUT charge. 
%The KTN spacetime is a stationary, axisymmetric vacuum solution of Einstein's equation with a Kerr parameter and NUT charge. With this, the contrast between the Kerr and the KTN spacetime is clearly realised. 
%Also, note 
%Note that both Kerr and KTN spacetimes are the vacuum solutions of Einstein's equation. 
Needless to mention that Kerr spacetime is described by mass and Kerr parameter ($a_{\rm k}$, the spin angular momentum per unit mass) of the black hole whereas three parameters are required to uniquely specify the KTN spacetime. These parameters are the mass, the Kerr parameter ($a_{\rm k}$) and the NUT parameter ($n$, also called NUT charge). In the limit $n\to 0$, the KTN spacetime reduces to the Kerr spacetime and if $a_{\rm k}\to 0$ the KTN spacetime reduces to the Taub-NUT spacetime. Finally, if the NUT parameter is also made to vanish in the Taub-NUT spacetime, it reduces to pure Schwarzschild solution. In the astrophysical context, although the Kerr spacetime is known to be very much relevant, however it is intriguing to explore KTN spacetime as well while studying the accretion flows around it. 

The KTN metric first appeared in the works of Demianski and Newman \cite{Demianski-Newman1966}. Later, this class of solution was derived and interpreted in different works by Carter \cite{Carter1966}, Kinnersley \cite{Kinnersley1969},  Kramer anad Neugebauer \cite{Kramer-Neugebauer1968},  Robinson et. al. \cite{Robinson-etal1969},  and Talbot \cite{Talbot-1969}. The observational 
possibilities of such spacetimes were first realized by Lynden-Bell and Nouri-Zonoz \cite{RevModPhys.70.427} in terms of shifting in spectral lines from quasars, supernovae, and active galactic nuclei. Few discrete studies 
were also carried out in order to understand the observational viability of such spacetimes \cite{Kagramanova_2010,Chakraborty_2019}. 

The interpretation of the NUT parameter has been the subject of debate specifically with regard to the existence of closed time like curve (CTC). The physical interpretation  that obviates the aforementioned pathological CTC was first suggested by Bonnor \cite{Bonnor1969}. The source of the metric is interpreted as a spherically symmetric mass together with a semi-infinite mass-less source of angular momentum along the symmetry axis.
In their original paper, Demianski \& Newman \cite{Demianski-Newman1966} interpreted the NUT parameter with a `dual mass' or gravitomagnetic monopole. Gravitomagnetic monopole is the gravitational analog of a magnetic monopole, which is often interpreted as a linear source of pure angular momentum \cite{Bonnor1969,Miller1972,Dowker1974,Ramaswamy-Sen1981}. 
The non-diagonal part of the metric featuring the NUT parameter ($n$) as a gravitomagnetic charge determines the gravitomagnetic properties of the Taub-NUT spacetime. 
The non-diagonal term implies a singularity
on the $\theta=\pi$ axis, which is known as the Misner string.  This type of singularity is completely different from the ordinary coordinate singularity. According to Misner \cite{Misner:1963fr}, this singularity can be avoided by introducing a periodic time
coordinate and two different coordinate patches. 
All these interpretations of the metric were abstract and theoretical in nature. 
However, because of the availability of the state-of-the-art astrophysical observation facilities now a days,
%However, given the improvement in the sophistication of direct observation methods in various astrophysical scenarios over the last decade, 
nothing could be more appropriate than taking a practical approach towards understanding the nature of those kind of non-standard gravitational backgrounds. 

Motivated by this, in the present work, we study the 
 accretion phenomena in a special class of gravitational background, namely the KTN spacetime. Since the accretion of matter around the compact object is believed to be the driving mechanism behind most of the energetic phenomena in the universe \cite{Frank-etal2002}, the study of the effect of the 
 NUT parameter on the accretion phenomena is expected to shed some light not only on the observational aspects but also on the theoretical understanding of the Einstein's theory itself. Keeping this goal in mind, in the present work, we consider an optically and geometrically thin accretion disc in the KTN background. 
 % to investigate the flow properties. 
% This would reveal the physical nature of the NUT parameter.  
In order to investigate the flow properties, we consider the full general relativistic hydrodynamic framework \cite{Rezzolla-Zanotti2013} with relativistic equation of state (EoS) \cite{Chandrasekhar1939, Synge1957, Cox_Giuli1968, Dihingia-etal2019a} . With these considerations, we perform the critical point analysis and obtain the solutions in terms of energy $({\cal E})$ and angular momentum $(\lambda)$ of the flow. Here, we choose $a_{\rm k}$ and $n$ values in such a way that the combination represents KTN spacetime. We then study the role of the NUT parameter in deciding the nature of the accretion solutions. 
Farther, we identify the parameter space in the $\lambda - {\cal E}$ plane for multiple critical points and examine how parameter space is modified with the increase of $a_{\rm k}$ and $n$, respectively.  
Moreover, we calculate the disc luminosities corresponding to the flow solutions characterized with a given set of ($a_{\rm k}$, $n$) and find the maximum luminosity $(L_{\rm max})$ in KTN spacetime. Finally, we extend our work for naked singularity as well, where we obtain the parameter space for multiple critical points and examine how the nature of the flow solution changes with $a_{\rm k}$ and $n$ values. 

We arrange the paper in the subsequent sections as follows. In section II \& III,  we formulate the mathematical building blocks to study the accretion flow and discuss the critical point analysis. In section IV, we describe the methodology to calculate the flow solutions and obtain the parameter space for multiple critical points. In sections V, we present the results obtained for naked singularity. Finally, in section VI, we present the discussion and conclusion of this work.

\section{Kerr-Taub-NUT (KTN) background}

%As already discussed in the introduction, we essentially 
We consider a special class of axisymmetric vacuum solution of Einstein's theory, which is the KTN metric expressed in Boyer-Lindquist coordinates as \cite{Carter1966},
$$\begin{aligned}
ds^2 = &g_{\mu\nu} dx^\mu dx^\nu,\\
= & g_{tt}dt^2 + 2g_{t\phi}dtd\phi + g_{rr} dr^2 + g_{\theta\theta} 
d\theta^2 + g_{\phi\phi} d\phi^2,
\end{aligned}\eqno(1)$$
where $x^\mu  (\equiv t,r,\theta,\phi)$ denote coordinates and
$g_{tt} = (a_{\rm k}^2 \sin ^2\theta-\Delta)/\Sigma $, $g_{t\phi} = (A \Delta - a_{\rm k} B \sin ^2\theta )/\Sigma$, $g_{rr}=\Sigma/\Delta$, $g_{\theta\theta}=\Sigma$ and $g_{\phi\phi}=(B^2\sin ^2\theta -A^2 \Delta )/\Sigma $ are the non-zero metric components. Here, $A=a_{\rm k} \sin ^2\theta-2 n \cos \theta $, $\Sigma = (a_{\rm k} \cos\theta+n)^2+r^2$, $B=a_{\rm k}^2+n^2+r^2$ and $\Delta = r^2 - 2r + a_{\rm k}^2 - n^2$. Throughout this paper, we adopt the sign convention as $(-,+,+,+)$. 
%We denote $a_{\rm k}$ as the Kerr parameter and $n$ as the NUT parameter of the metric, respectively. 
We set the source mass $M_{\rm S}=1$, and work in units where the universal gravitational constant $G=1$ and the speed of light $c=1$ is used. In this unit system, we express length, angular momentum, and time in terms of $GM_{\rm S}/c^2$, $GM_{\rm S}/c$ and $GM_{\rm S}/c^3$, respectively.

The event horizon ($r_{\rm H}$) of the metric is defined as $\Delta=0$, which gives
$$
r_{\rm H}=1+ \sqrt{1-a_{\rm k}^2+n^2}.
\eqno(2)
$$
We have taken only the outer horizon as the region of our interest. Equation (2) clearly suggests
that depending on the values of $a_{\rm k}$ and $n$, KTN spacetime represents either black hole with $(1-a_{\rm k}^2+n^2) >0$ or naked singularity with ($1-a_{\rm k}^2+n^2)<0$. For $n=0$, KTN spacetime boils down to the usual Kerr spacetime. One of the most important physical implications of this background, contrary to the conventional wisdom, is that the spin parameter $a_{\rm k}$ can now be larger than unity for black holes. This particular fact specifically makes KTN black hole spacetime a fertile ground for accretion study having much richer phenomenology as compared to usual Kerr black holes. 

\section{Assumptions and Model Equations}

In this paper, we carry out the hydrodynamical analysis of accretion flow based on some simple set of assumptions. We consider the flow to be axisymmetric in accordance with the KTN background. For simplicity, we also consider non-dissipative, optically, and geometrically thin accretion flow in the steady state \cite{Chakrabarti1989,Chakrabarti1996,Das-etal2001}.

\subsection{Equations of the fluid}

In the framework of the relativistic accretion processes, the non-dissipative energy-momentum tensor for the fluid composed of ions and electrons can generally be expressed as, 
$$
T^{\mu\nu} = (e+p)u^\mu u^\nu + pg^{\mu\nu},
\eqno(3)
$$
where $e$, $p$, and $u^\mu$ represent the energy density, pressure, and the four velocities of any fluid element, respectively. The time-like velocity field satisfies the following local normalization condition $u_\mu u^\mu=-1$. Here, $\mu$ and $\nu$ are the spacetime indices running from $0 \to 3$, and $g^{\mu\nu}$ are the components of the metric under consideration. The conservation of energy-momentum tensor and the mass flux give all the hydrodynamical equations required to describe the flow, and 
%They 
are given by,
$$
T^{\mu\nu}_{;\nu}=0,\qquad (\rho u^\nu)_{;\nu}=0,
\eqno(4)
$$ 
where $\rho$ is the local mass density of the flow. In relativistic hydrodynamics, we employ the projection operator  $h^i_\mu = \delta^i_\mu + u^i u_\mu$, where  `$i$' takes only the spatial coordinates, which satisfies $h^i_\mu u^{\mu}= 0$. By taking the projection of conservation equation on the spatial hypersurface, one obtains the relativistic Euler equation,
$$
h^i_\mu {T^{\mu\nu}}_{;\nu}=(e+p)u^\nu u^i_{;\nu} + (g^{i\nu} + u^i u^\nu)p_{,\nu}=0 , 
\eqno(5)
$$
and, projecting it along $u^{\mu}$, we have the first law of thermodynamics as, 
$$
u_\mu T^{\mu\nu}_{;\nu}=u^\mu\bigg[\left(\frac{e+p}{\rho}\right)\rho_{,\mu} - e_{,\mu}\bigg]=0.
\eqno(6)
$$

To describe the flow completely, we need to know the equation of state (EoS) of the fluid under consideration, which relates the density $(\rho)$, pressure $(p)$, and the internal energy $(e)$ of the flow. Usually, the temperature of the accretion flow  
can go up to $\sim 10^{10-11}$K \cite{Dihingia-etal2018b}, particularly, within a  few Schwarzschild radius. Therefore, we consider the relativistic EoS given by Chattopadhyay \& Ryu \cite{Chattopadhyay-Ryu2009}, 
$$
e = n_em_ef=\frac{\rho}{\tau}f.
\eqno(7)
$$
Here, $n_e$ and $m_e$ are the number density and the mass of the electrons, $\rho = n_em_e\tau$, $\tau = [2 - \zeta(1 - 1/\chi)]$, $\zeta = n_p/n_e$, and $\chi=m_e/m_p$, respectively, where  $n_p$ and $m_p$ are the number density and the mass of the ions. We consider the flow to be composed of solely by ions and electrons. Hence, throughout our study, we set $\zeta=1$, until otherwise stated. Finally, the extended form of $f$ in terms of the dimensionless temperature $\Theta~(= k_{\rm B}T/m_ec^2)$  is given by,
$$
f = (2-\zeta)\bigg[1 + \Theta\left(\frac{9\Theta + 3}{3\Theta + 2}\right)\bigg] +
 \zeta\bigg[ \frac{1}{\chi} + \Theta\left(\frac{9\Theta + 3/\chi}
{3\Theta + 2/\chi}\right)\bigg].
\eqno(8)
$$
According to the relativistic EoS, the explicit expressions of the polytropic index $(N)$, adiabatic index $(\Gamma)$ and the sound 
speed $(a_s)$ are given as,
$$
N = \frac{1}{2}\frac{df}{d\Theta}; \quad \Gamma = 1 + \frac{1}{N}; {\rm and}
\quad a_s^2 = \frac{\Gamma p}{e+p} = \frac{2\Gamma\Theta}{f + 2\Theta}.
\eqno(9)
$$
In equation (9), $N$ and $\Gamma$ are expressed as a function of $\Theta$, and therefore, these quantities would be determined self consistently while obtaining the flow properties across the length-scale of the accretion disk.
	
\subsection{Governing Equations for Accretion Disc}

In our analysis, we assume geometrically thin accretion disc around black hole in the steady state. Therefore, given the background axisymmetry, one can generically consider the disc to be lying on the equatorial plane with $\theta=\pi/2$, and consequently $u^\theta\sim 0$. Under this assumption, the radial component of the relativistic Euler equation (equation (5)) takes the following form,
$$
\begin{aligned}
&u^ru^r_{,r} + \frac{1}{2}g^{rr}\frac{g_{tt,r}}{g_{tt}} + \frac{1}{2}u^ru^r
\left(\frac{g_{tt,r}}{g_{tt}} + g^{rr}g_{rr,r}\right)\\
 + &u^\phi u^tg^{rr}\left(\frac{g_{t\phi}}{g_{tt}}
g_{tt,r} - g_{t\phi,r}\right)
+\frac{1}{2}u^{\phi}u^{\phi}g^{rr}\left(\frac{g_{\phi\phi}g_{tt,r}}{g_{tt}} - g_{\phi\phi,r}\right)\\
&+ \frac{(g^{rr} + u^ru^r)}{e+p}p_{,r}=0.\\
\end{aligned}
\eqno(10)
$$
Subsequently, the second part of the equations (4), $i.e$., the continuity equation can be expressed in terms of the mass accretion rate $(\dot{M})$, which is a constant of motion and is given by,
$$
\dot{M} = -4\pi r u^r \rho H,
\eqno(11)
$$
where $H$ is the local half-thickness of the accretion disc. The functional form of $H$ is obtained by following Riffert \& Herold \cite{Riffert-Herold1995}, and Peitz \& Appl \cite{Peitz-Appl1997}, in the form,
$$
H^2 = \frac{pr^3}{\rho \mathcal{F}},
\eqno(12)
$$
with
$$
\mathcal{F}=\gamma_\phi^2\frac{(r^2 + a_k^2)^2 + 2\Delta a_k^2}
{(r^2 + a_k^2)^2 - 2\Delta a_k^2},
$$
where $\gamma_\phi^2=1/(1-v_\phi^2)$ and $v_\phi^2 = u^\phi u_\phi/(-u^t u_t)$, respectively. It has been shown to be convenient and physically transparent to study the dynamics in terms of all the flow variable defined in the co-rotating frame \citep{Chakrabarti1996,Peitz-Appl1997}. 
In the co-rotating frame, the radial three velocity is defined as $v^2 = \gamma_\phi^2v_r^2$ and thus the associated radial Lorentz factor is given by $\gamma_v^2 = 1/(1-v^2)$, where $v_r^2 = u^ru_r/(-u^tu_t)$. Employing these definitions of the velocities and using the expressions $g^{\mu \nu}$ for KTN metric in equation (10), we obtain
$$
v\gamma_v^2\frac{dv}{dr} + \frac{1}{h\rho}\frac{dp}{dr} + \frac{d\Phi^{\rm eff}_e}{dr}=0,
\eqno(13)
$$
where $h\left[=(e+p)/\rho\right]$ is the specific enthalpy, $\Phi^{\rm eff}_e$ denotes the effective potential at the disc equatorial plane, and is given by \citep{Dihingia-etal2018a},
$$
\Phi^{\rm eff}_e=1+\frac{1}{2} \ln \left(\frac{\left(n^2+r^2\right) \Delta }{\left(a_{\rm k}^2-a_{\rm k} \lambda +n^2+r^2\right)^2-(a_{\rm k}-\lambda )^2 \Delta}\right).
\eqno(14)
$$

\begin{figure}
	\includegraphics[scale=0.4]{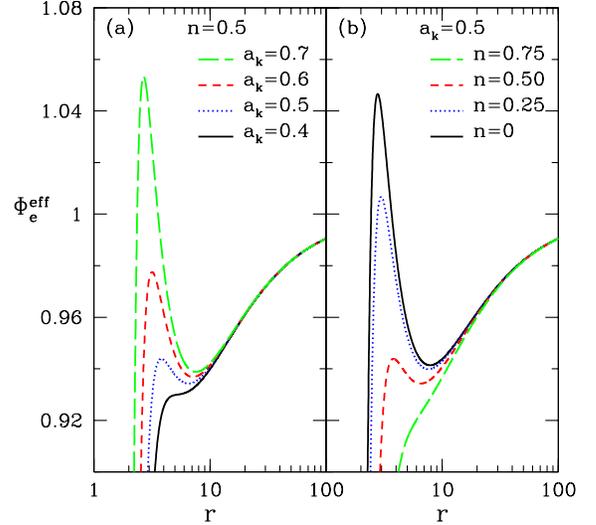}
	\caption{Plot of effective potential $\Phi^{\rm eff}_{\rm e}$ with radial distance for angular momentum $\lambda = 3.5$. In the left panel, we fix NUT parameter $n=0.5$, and solid (black), dotted (blue), small-dash (red) and long-dash (green) curves are for Kerr parameter $a_{\rm k} = 0.4, 0.5, 0.6$ and $0.7$, respectively. In the right panel, we fix $a_{\rm k}=0.5$, and solid (black), dotted (blue), small-dash (red) and long-dash (green) curves are for $n = 0, 0.25, 0.5$ and $0.75$, respectively.
		}
	\label{fig:r_phi_ak_n}
\end{figure}

In Fig. \ref{fig:r_phi_ak_n}, we illustrate the variation of effective potential ($\Phi^{\rm eff}_e$) with radial coordinate ($r$) for a given angular momentum $\lambda = 3.5$. In the left panel of the figure, we demonstrate how $\Phi^{\rm eff}_e$ varies with black hole spin ($a_{\rm k}$). Here, we choose NUT parameter $n=0.5$, and solid (black), dotted (blue), small-dash (red) and long-dash (green) curves are for Kerr parameter $a_{\rm k} = 0.4, 0.5, 0.6$ and $0.7$, respectively. Similarly, in the right panel of the figure, we show the dependencies of $\Phi^{\rm eff}_e$ on $n$, where $a_{\rm k} =0.5$ is used. Here, solid (black), dotted (blue), small-dash (red) and long-dash (green) curves are obtained for $n = 0, 0.25, 0.5$ and $0.75$, respectively. Figure \ref{fig:r_phi_ak_n} clearly indicates that for KTN black hole spacetime, the role of $a_{\rm k}$ and $n$ are opposite to each other in deciding the features of $\Phi^{\rm eff}_e$.

Because the KTN spacetime is stationary and axisymmetric, there exists two mutually perpendicular Killing vectors, namely $\partial_t$ and $\partial_\phi$. These two Killing vectors helps to construct two conserve quantities along the direction of the motion, and are given by,
$$
hu_\phi = {\cal L} ~({\rm constant}); \qquad -hu_t={\cal E} ~({\rm constant}),
\eqno(15)
$$
where 
${\cal E}$ is
the Bernoulli constant (equivalently specific energy) of the flow. Here, $u_t = -\gamma_v \gamma_\phi/\sqrt{\lambda g^{t\phi} - g^{tt}}$, where $\lambda = -u_\phi/u_t$ is the specific angular momentum of the flow which is also a constant of motion obvious from equation (15). 

Integrating equation (6) with the help of equation (7-8), we obtain the expression of density ($\rho$) in terms of temperature ($\Theta$) as,
$$
\rho={\cal K}\exp(k_3)\Theta^{3/2}(3\Theta +2)^{k_1}(3\Theta + 2/\chi)^{k_2},
\eqno(16)
$$
where ${\cal K}$ is the entropy constant and $k_1=3(2-\zeta)/4$, $k_2=3\zeta/4$, and $k_3=(f-\tau)/(2\Theta)$. Following \citet{Chattopadhyay-Kumar2016,Kumar-Chattopadhyay2017}), we define the entropy accretion rate ($\dot {\cal M}$) as
$$
\dot{\cal{M}}=\frac{\dot{M}}{4\pi {\cal K}}=\exp(k_3)\Theta^{3/2}(3\Theta +2)^{k_1}(3\Theta + 2/\chi)^{k_2}Hru^r.
\eqno(17)
$$
It maybe noted that $\dot {\cal M}$ is also a constant of motion.

\subsection{Wind Equation}

To obtain the wind equation, we make use of equations (6), (7), (10), and (11).
In fact, it is customary to express the wind equation as follows, 
$$
\frac{dv}{dr}= \frac{\mathcal{N}}{\mathcal{D}},
\eqno(18)
$$
where denominator $\mathcal{D}$ is given by,
$$
\mathcal{D} = \gamma_{v}^2\bigg[v- \frac{2a_s^2}{v(\Gamma +1)}\bigg],
\eqno(19)
$$
and numerator $\mathcal{N}$ is given by,
$$
\mathcal{N}=\frac{2a_s^2}{\Gamma + 1}\bigg[ \frac{1}{2\Delta}\frac{d\Delta}{dr} + \frac{1}{2\eta}\frac{d\eta}{dr}+ \frac{3}{2r} - \frac{1}{2\mathcal{F}}\frac{d\mathcal{F}}{dr}\bigg] - \frac{d\Phi^{\rm eff}_e}{dr}.
\eqno(20)
$$
Here, we write $\eta=r^2/(r^2+n^2)$.

Similarly, the gradient of the temperature is obtain by rewriting equation (5) using equations (9) and(11) as,
$$
\frac{d\Theta}{dr}=-\frac{2\Theta}{2N + 1}\bigg[\frac{\gamma_{v}^2}{v}\frac{dv}{dr} - \frac{1}{2\mathcal{F}}\frac{d\mathcal{F}}{dr} +\frac{1}{2\Delta}\frac{d\Delta}{dr}+\frac{1}{2\eta}\frac{d\eta}{dr}+\frac{3}{2r}\bigg].
\eqno(21)
$$ 

\subsection{Critical Point analysis}

In order to obtain the accretion solution, one requires to solve equations (18) and (21) simultaneously by using the initial condition of the flow. In this work, the initial condition of the flow is characterized by a set of input parameters, namely the radial velocity $v(r)$, temperature $\Theta(r)$ and the angular momentum $(\lambda)$ of the flow along with the spacetime parameters $(a_{\rm k},n)$. Interestingly, because of the nature of the black hole spacetime, the accretion flow around a black hole must be transonic, which essentially means that while accreting towards the black hole, the flow must make smooth transition from sub-sonic to supersonic velocity at some point before entering the black hole. Such a special point where flow changes its sonic character is called as critical point ($r_{\rm c}$). At the critical point, the numerator and the denominator of the wind equation (18) vanish simultaneously ($i.e., dv/dr=0/0$) where we have the critical point conditions as ${\cal N}={\cal D}=0$. To calculate the radial velocity gradient $(dv/dr)_{\rm c}$ at $r_{\rm c}$, we apply the l$^{\prime}$Hospital rule. In general, $(dv/dr)_{\rm c}$ possesses two distinct values; one of them is for accretion and the other one is for wind. When both values of $(dv/dr)_{\rm c}$ are real and of opposite sign, the critical point is called as saddle type critical point; when $(dv/dr)_{\rm c}$ are real and same sign it is called as nodal type critical point (or N-type) and when $(dv/dr)_{\rm c}$ are complex, it is the spiral type (or O-type) critical point \citep[and references therein]{Chakrabarti-Das2004}. It may be noted that saddle type critical points have special importance as the global transonic accretion flow can only pass through it. 
In reality, depending on the input parameters, the flow may possess single or multiple critical points within the length scale of the accretion disc \cite{Fukue1987,Chakrabarti1989}. When critical points form close to the horizon, it is referred as the inner critical points ($r_{\rm in}$) and when they form far away from the black hole, it is termed as the outer critical points ($r_{\rm out}$). 

\section{accretion around KTN black hole}

In this section, we intend to focus on the KTN black hole background keeping the naked singularity case aside for discussion in Section V. In reality, the behavior of the accretion flow depends on both the parameters describing the KTN black hole, namely Kerr ($a_{\rm k}$) and NUT parameters ($n$). However, since the role of $a_{\rm k}$ in studying the accretion solution around black hole is already well explored, we plan is to concentrate on the NUT parameter ($n$) only and investigate its impact on the flow properties.

\subsection{Properties of the critical points}

\begin{figure}
	\includegraphics[scale=0.4]{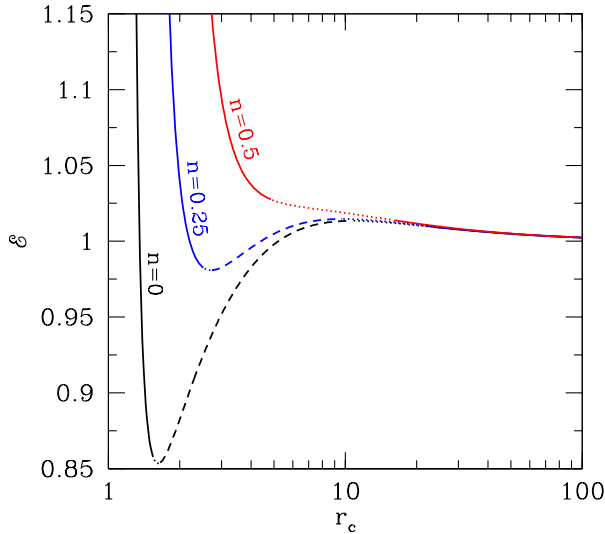}
	\caption{Plot of energy as a function of critical point location ($r_c$) for three different NUT parameter, namely $n=0$, $n=0.25$, and $n=0.5$. Here, we choose $a_{\rm k}=0.99$, and $\lambda=2.04$. Solid , dotted and dashed curves denote saddle, nodal and O-type critical points, respectively. See text for details.}
	\label{fig:rc_E_n}
\end{figure}

Since the accretion solutions around the black holes can only pass through the saddle type critical points, it is useful to examine how the nature of the critical points depends on the NUT parameters ($n$). For that we choose a set of ($\lambda, a_{\rm k}, n$) values to calculate the flow energy (${\cal E}$) at the critical points by using the critical point conditions. The obtained results are shown in Fig. \ref{fig:rc_E_n} where we plot the variation of ${\cal E}$ with the critical point location ($r_c$) for different values of $n$. Here, we choose $ a_{\rm k}=0.99 $ and $\lambda=2.04$. In the figure, the left to right curves are obtained for $n=0$, $0.25$ and $0.5$ respectively and $n$ values are marked. In a given curve, we denote the saddle, nodal and O-type critical points by solid, dotted and dashed line styles, respectively. For $n=0$, it is observed that the nature of the critical points changes in systematic order as saddle$-$nodal$-$spiral$-$nodal$-$saddle with the shift of the location of the critical points from the black hole. We also observe that there exists an energy range that allows the flow to possess maximum of three critical points. Out of the three critical points, one is O-type, and the other two are either saddle type or combination of spiral and nodal types. For ${\cal E} <1$, flow contains two critical points, and between them one is saddle or nodal type and other is O-type. When energy is above a critical value, flow only owns a single saddle type critical point located close to the horizon. For $n=0.25$, we find similar results as in the case of $n=0$, however, the energy range for multiple critical points is reduced and the locations of the critical points are shifted outwards for a given energy. When NUT parameter is increased further to $n=0.5$, we observe that multiple critical points completely disappear. This clearly indicates that there exists a critical NUT parameter (say, $n^{\rm cri}$) beyond which multiple critical points do not exist. Needless to mention that $n^{\rm cri}$ does not have any universal value, instead it strongly depends on the other input parameters. Overall, the above analysis suggests that NUT parameter ($n$) plays an important role in deciding the properties of the accretion flow in KTN black hole background.

\subsection{Global Transonic Accretion solution}

In order to solve the hydrodynamic equations, the aforementioned criticality condition plays very important role in identifying the appropriate boundary conditions for the flow. Setting ${\cal D}=0$ and ${\cal N}=0$, and using $ {\cal E}=-hu_t $ (see equation (15)), we obtain the radial velocity ($v_{\rm c}$) and temperature $(\Theta_{\rm c})$ at the critical point $(r_{\rm c})$ for a given set of (${\cal E}, \lambda, a_{\rm k}, n$) values. In other words, these two critical point conditions enable us to reduce the number of input parameters from six to four and therefore, we can start the integration of equations (18) and (21) from the critical point itself to obtain the global accretion solutions.
Accordingly, using the same set of (${\cal E}, \lambda, a_{\rm k}, n$) values as the input parameters, we integrate equation (18) and (21) from the critical point ($r_{\rm c}$) first up to horizon and then up to a large distance (equivalently the outer edge of the disc). Finally, we join these two parts of the solution to obtain a global transonic accretion solution around black hole \cite[and references therein]{Dihingia-etal2018a,Dihingia-etal2019a,Dihingia-etal2019b}.
 
 \begin{figure*}
 	\vskip -3.5cm
 	\includegraphics[scale=0.7]{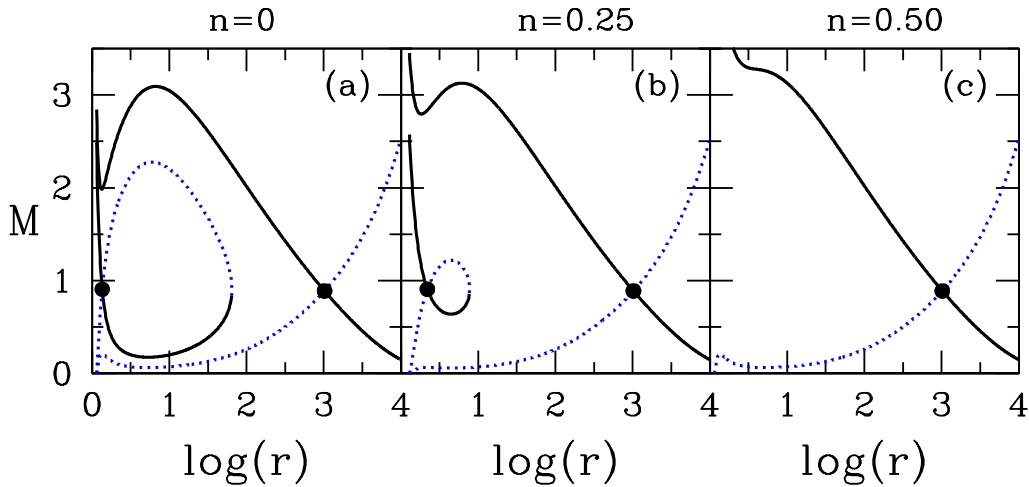}
 	\vskip -4cm
 	\caption{Plot of Mach number $(M=v/a_s)$ as a function of radial distance ($r$). Solid (black) curve represents accretion solution and dotted (blue) curve denotes winds. Here, we choose ${\cal E}=0.0001$, $\lambda = 2.04$, and $a_{\rm k} = 0.99$ for all panels. Results presented in panel (a), (b) and (c) are for $n=0$, $0.25$ and $0.5$, respectively. See text for details. 
 		}
 	\label{fig:r_M_n}
 \end{figure*}

 Following the above procedure, we calculate the global flow solutions for different $n$ values and plot them in Fig. \ref{fig:r_M_n}. While obtaining the solution, we fix ${\cal E}=1.0001$, $\lambda = 2.04$, and $a_{\rm k}=0.99$, and vary NUT parameter as in panel (a) $n=0$, (b) $0.25$, and (c) $0.5$. In each panel, the Mach number ($M=v/a_s$) is plotted as function of radial coordinate where solid (black) and dotted (blue) curves denote accretion and wind branches, respectively and filled circles represent the critical points. Figure clearly shows how the nature of the flow solutions changes with $n$ for a given set of (${\cal E}, \lambda, a_{\rm k}$) values. In panel (a), the flow possesses multiple critical points, and the solution passing through the inner critical point $(r_{\rm in}=1.3542)$ is closed and fails to connect the event horizon to the outer edge of the accretion disc. On the other hand, the solution passing through the outer critical point $(r_{\rm out}=1021.9821)$ smoothly connects the event horizon to the outer edge of the accretion disc. Interestingly, after crossing the outer critical point, accretion solution may join with the solution passing through the inner critical point via shock transition because of the fact that the latter solution possesses higher entropy \cite[references therein]{Das2007}. Accretion solutions of this kind are potentially promising in the astrophysical context and will be reported elsewhere. 
 When NUT parameter is increased to $n=0.25$ (see panel (b)), flow continues to possess multiple critical points ($r_{\rm in}=2.2010,r_{\rm out}=1022.4793$), however, the closed solution passing through the inner critical point terminates at smaller radii. For further increase of NUT parameter ($n =0.5$), the inner critical point vanishes and solution has the only option to pass through the outer critical point $(r_{\rm out}=1023.9684)$ only.

\subsection{Parameter space with NUT charge}

\begin{figure}
	\includegraphics[scale=0.4]{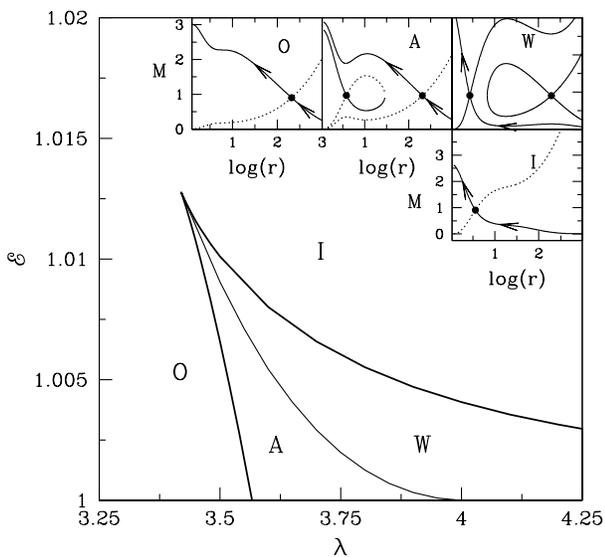}
	\caption{Region of the parameter space in $\lambda-{\cal E}$ plane according to the nature of the flow solutions. At the insets, all possible flow solutions (O, A, W, I) are presented. See text for details.
	}
	\label{fig:L_E_r_M}
\end{figure}

In this section, we begin with the study of parameter space in $\lambda-{\cal E}$ plane according to the nature of flow solutions.  
To do that we fix $a_{\rm k} = 2.23$ and $n = 2$ and vary both $\lambda$ and ${\cal E}$ freely to calculate various flow solutions. Here, we restrict our investigation for ${\cal E} \ge 1$ as bounded energy does not provide complete global accretion solutions \cite{Chakrabarti1996}. All together, four different types of flow solutions are found and accordingly, four distinct regions of the parameter space are identified which are marked as `O', `A', `W' and `I', respectively, in Fig. \ref{fig:L_E_r_M}. 
The representative flow solutions from these regions are depicted at the inset panels where the variation of Mach number ($M=v/a_s$) with radial coordinate is shown and individual panels are also marked. In each panel, solid curves denote the accretion solutions, dotted curves indicate the wind solutions and filled circles represent the critical points. Arrows indicate the overall direction of flow motion towards the black hole.

\begin{figure}
	\includegraphics[scale=0.4]{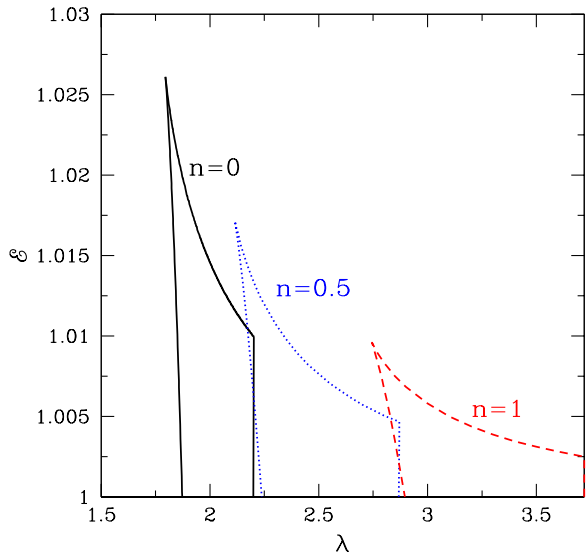}
	\caption{Modification of the parameter space for multiple critical points with the increase of NUT parameter ($n$). Region bounded with solid (black), dotted (blue) and dashed (red) are for $n=0.0, 0.5$, and $1.0$, respectively. Here, we fixed the Kerr parameter $a_{\rm k}=0.99$. See text for details.
	}
	\label{fig:L_E_n}
\end{figure}

Next, we examine the range of flow parameters that provides the flow solutions containing multiple critical points around the black hole having spin $a_{\rm k}=0.99$.
 While doing this, we fix NUT parameter and vary the remaining parameters, namely ${\cal E}$ and $\lambda$ freely. This allows us to obtain the parameter space spanned by ${\cal E}$ and $\lambda$ which is depicted in Fig. \ref{fig:L_E_n}. Effective region of the parameter space separated by solid (black), dotted (blue) and dashed (red) boundaries are obtained for $n=0$, $0.5$ and $1$, respectively which are marked in the figure. It is clear from the figure that as NUT parameter is increased, the accretion flow harbors multiple critical points provided its energy is lower and angular momentum is higher. 

\begin{figure}
	\includegraphics[scale=0.4]{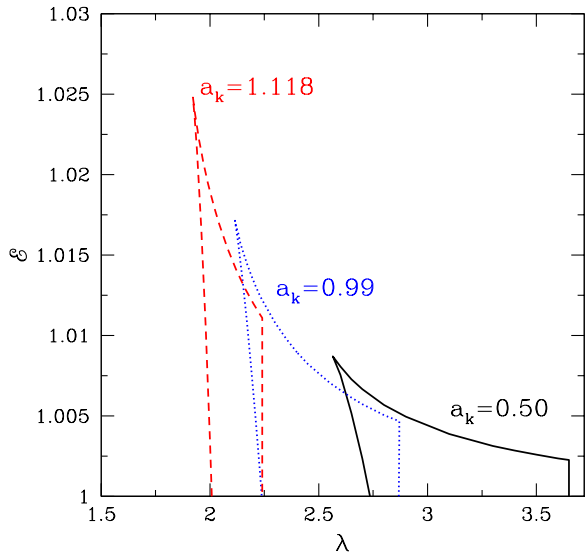}
	\caption{Same as Fig. \ref{fig:L_E_n}, but NUT parameter is fixed as $n=0.5$ and $a_{\rm k}$ is varied as marked in the figure.
		}
	\label{fig:L_E_ak}
\end{figure}

In Fig. \ref{fig:L_E_ak}, we depict the modification of parameter space for multiple critical points as $a_{\rm k}$ is varied. Here, we fix the NUT parameter as $n=0.5$ and the region bounded using solid (black), dotted (blue) and dashed (red) curves are calculated for $a_{\rm k}=0.5, 0.99$ and $1.118$, respectively. We observe that as $a_{\rm k}$ is increased keeping $n$ fixed, the parameter space for multiple critical points is shifted toward higher energy and lower angular momentum domain. By comparing Fig. \ref{fig:L_E_n} and Fig. \ref{fig:L_E_ak}, we also find that $n$ and $a_{\rm k}$ play opposite role in determining the parameter space as expected.

\begin{figure}
	\includegraphics[scale=0.4]{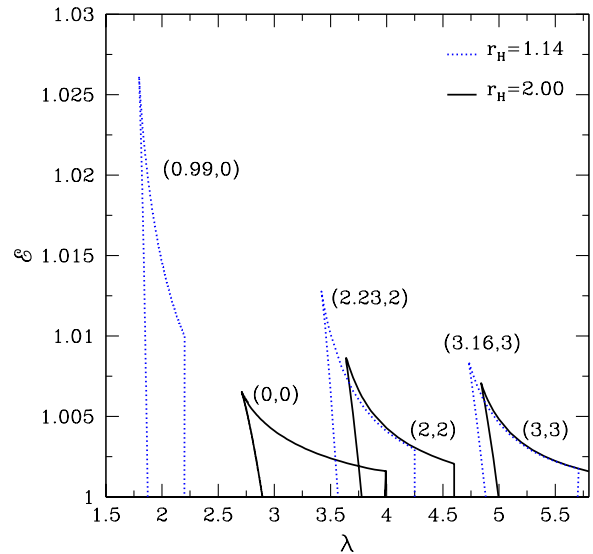}
	\caption{Comparison of parameter space for multiple critical points. Effective regions of the parameter space bounded with dotted (blue) and solid (black) curves are obtained for two different event horizon locations $r_{\rm H}=1.14$ and $r_{\rm H}=2$, respectively. The Kerr parameter and NUT parameter ($a_{\rm k},n$) are marked in the figure. see text for details.
	}
	\label{fig:L_E_ak_n}
\end{figure}

More interesting and rich phenomenology comes into play when the fast spinning KTN black holes are taken into considerations. It has already been emphasized that KTN black hole can accommodate spin parameter $a_{\rm k}$ larger than unity as opposed to the usual Kerr black hole provided the chosen NUT parameter satisfies the condition $1-a_{\rm k}^2+n^2>0$. Therefore, KTN spacetime opens up new opportunities to explore large class of observations which may not be possible in usual 
Kerr black hole spacetime (for a recent work see \cite{Chandrachur-Bhattacharyya2018}). Keeping this in mind, we study the accretion flow dynamics around the rotating black holes with Kerr parameter $a_{\rm k}>1$, and compare the results with that of the usual Kerr black hole $(n=0)$. In order to do so, we choose different combination of Kerr and NUT parameters keeping the event horizon location ($r_{\rm H}$) fixed and identify the ranges of ${\cal E}$ and $\lambda$ that admit accretion solutions containing multiple critical points. The obtained results are depicted in Fig. \ref{fig:L_E_ak_n}, where we identify the region of parameter space in $\lambda-{\cal E}$ plane that render multiple critical points. Here, solid and dotted boundaries refer to the two event horizon locations as $r_{\rm H}=2$ (solid) and $r_{\rm H}=1.14$ (dotted). The chosen set of $(a_{\rm k},n)$ values are marked in the figure. We notice that for a given $n$, as $a_{\rm k}$ is increased, accretion flow generally possesses multiple critical points at lower angular momentum and higher energy ranges. Furthermore, $n$ and $a_{\rm k}$ play competing role in deciding the black hole horizon (see equation (2)) for KTN spacetime. Therefore, when $n \gg 1$, $r_{\rm H}$ tends to be insensitive to $a_{\rm k}$ causing the parameter space for multiple critical points indistinguishable as seen in Fig. \ref{fig:L_E_ak_n}.

\subsection{Radiative properties in KTN spacetime}

\begin{figure}
	\includegraphics[scale=0.4]{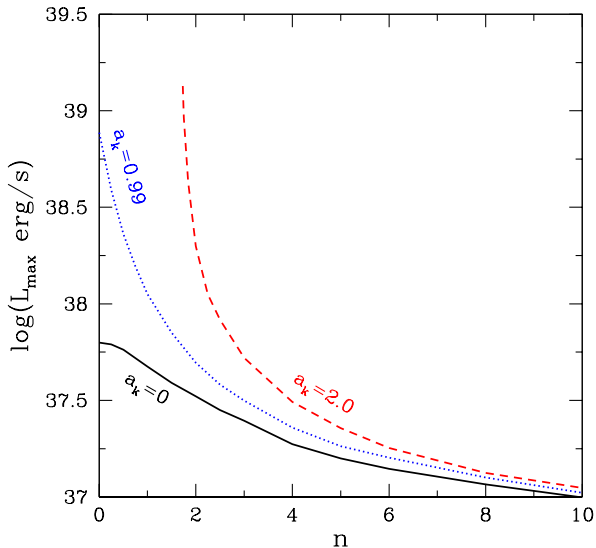}
	\caption{Plot of maximum luminosity $(L_{\rm max})$ as a function of NUT parameter $(n)$. Solid (black), dotted (blue) and dashed (red) curves denote the results corresponding to $a_{\rm k}=0.0, 0.99$ and $2.0$, respectively. Here, we choose black hole mass $M_{\rm S}=10~M_\odot$, accretion rate $\dot{m}=0.1~{\dot M}_{\rm Edd}$, and inclination angle $i=\pi/4$.}
	\label{fig:n_Lum_ak}
\end{figure}

During the course of accretion, inflowing matter experiences compression that causes the flow to become hot and dense. Hence, the flow is expected to emit high energy radiation. Since accretion flow is composed of both ions and electrons, free-free emission is viable and therefore, we consider the bremsstrahlung radiation from the accretion disc. Usually, the bremsstrahlung emission rate 
per unit volume, per unit time, per unit frequency 
is estimated as \cite{Vietri2008},
$$
\epsilon(\nu)=\frac{32 \pi e^6}{3 m_e c^2}\left(\frac{2\pi}{3k_{\rm B}m_e T_e}\right)^{1/2}n_e^2 e^{-h\nu/k_{\rm B}T_e}g_{\rm br},
\eqno(22)
$$
where $e$ is the charge of the electron, $h$ is the Planck's constant, $T_e$ is the electron temperature, $\nu$ is the frequency and $g_{\rm br}$ is the Gaunt factor. Note that $g_{\rm br}$ is a dimensionless quantity that varies between $0.2$ to $5.5$ \cite{Karzas1961} and in this work, we consider $g_{br}=1$ for simplicity. Further, following the work of \citet{Chattopadhyay-Chakrabarti2002}, we estimate the electron temperature ($T_e$) as $T_e=\left({m_e}/{m_p}\right)^{1/2}T$, where ${m_e}$ and ${m_p}$ denote the electron and ions masses, and $T$ refers the flow temperature. The total luminosity emitted from the accretion disc is obtained upon integrating $\epsilon(\nu)$ over the total volume and is given by 
$$
L=2\int_0^\infty \int_{r_{\rm H}}^{r_{\rm edge}}\int_0^{2\pi}Hr\epsilon(\nu_{\rm e})d\nu_{\rm e} dr d\phi.
\eqno(23)
$$
Here, $\nu_{\rm e}$ refers the emitted frequency which is related to the observed frequency $(\nu_{\rm o})$ as $\nu_{\rm e}=(1+z)\nu_{\rm o}$, where $z$ denotes the red-shift factor. Following \citet{Luminet1979}, we obtain $z$ as
$$
1+z=u^t(1+r \Omega \sin\phi\sin i),
\eqno(24)
$$
where $\Omega=u^\phi/u^t$ is the angular velocity and $i$ is the inclination angle of the black hole. In this work, we consider $i=\pi/4$ all throughout. Furthermore, we choose the black hole mass $M_{\rm S}=10 ~M_{\odot}$, where $M_{\odot}$ is the solar mass and accretion rate $\dot{m}=0.1~ {\dot M}_{\rm Edd}$,  ${\dot M}_{\rm Edd}$ being Eddington accretion rate. With this set up, we calculate the maximum disc luminosity $(L_{\rm max})$.
%and in Fig. \ref{fig:n_Lum_ak}, we depict the variation of $L_{\rm max}$ as a function of NUT parameter $(n)$ for different Kerr parameters.
%In order to calculate $L_{\rm max}$, 
While doing so, we choose a set of ($a_{\rm k}, n$) values and freely vary the energy $({\cal E})$ and angular momentum $(\lambda)$ of the flow. This provides the swarm of transonic accretion solutions [each solution is obtained for a particular set of ($a_{\rm k}, n, {\cal E}, \lambda$)] that are used in equation (23) to calculate the disc luminosity ($L$). Upon comparing various $L$ values, we find the maximum disc luminosity $(L_{\rm max})$. It is noteworthy to mention that during integration, we truncate the accretion disc at the outer edge $(r_{\rm edge})$ where $H/r \rightarrow 0.8$. 
The obtained results aref shown in Fig. \ref{fig:n_Lum_ak}, where the variation of $L_{\rm max}$ is plotted as a function of NUT parameter $(n)$ for different $a_{\rm k}$ values. In the figure, solid, dotted and dashed curves denote the results obtained for $a_{\rm k}=0$ (black), $a_{\rm k}=0.99$ (blue), and $a_{\rm k}=2.0$ (red), respectively. Note that for pure Kerr black hole $(n=0)$, $L_{\rm max}$ increases with the increase of $a_{\rm k}$. On the other hand, as $n$ gradually increases, $L_{\rm max}$ decreases. Moreover, we find that for Schwarzschild type KTN black hole $(a_{\rm k}=0)$, the rate of decrease of $L_{\rm max}$ with $n$ is relatively smaller compared to that of the rotating KTN black hole. Overall, we observe that $L_{\rm max}$ generally decreases with increasing $n$ irrespective of $a_{\rm k}$ values and appears to merge for large NUT parameter.

\section{Hydrodynamical flow around KTN naked singularity}

In this section, we study the properties of the hydrodynamic flow around the KTN naked singularity. The main motivation here is to explore the role of $a_{\rm k}$ and $n$ in deciding the nature of the critical points and the flow solutions. To do that we follow the same methodologies as discussed in Section IV.

\subsection{Properties of critical points}

\begin{figure*}
	\includegraphics[scale=0.4]{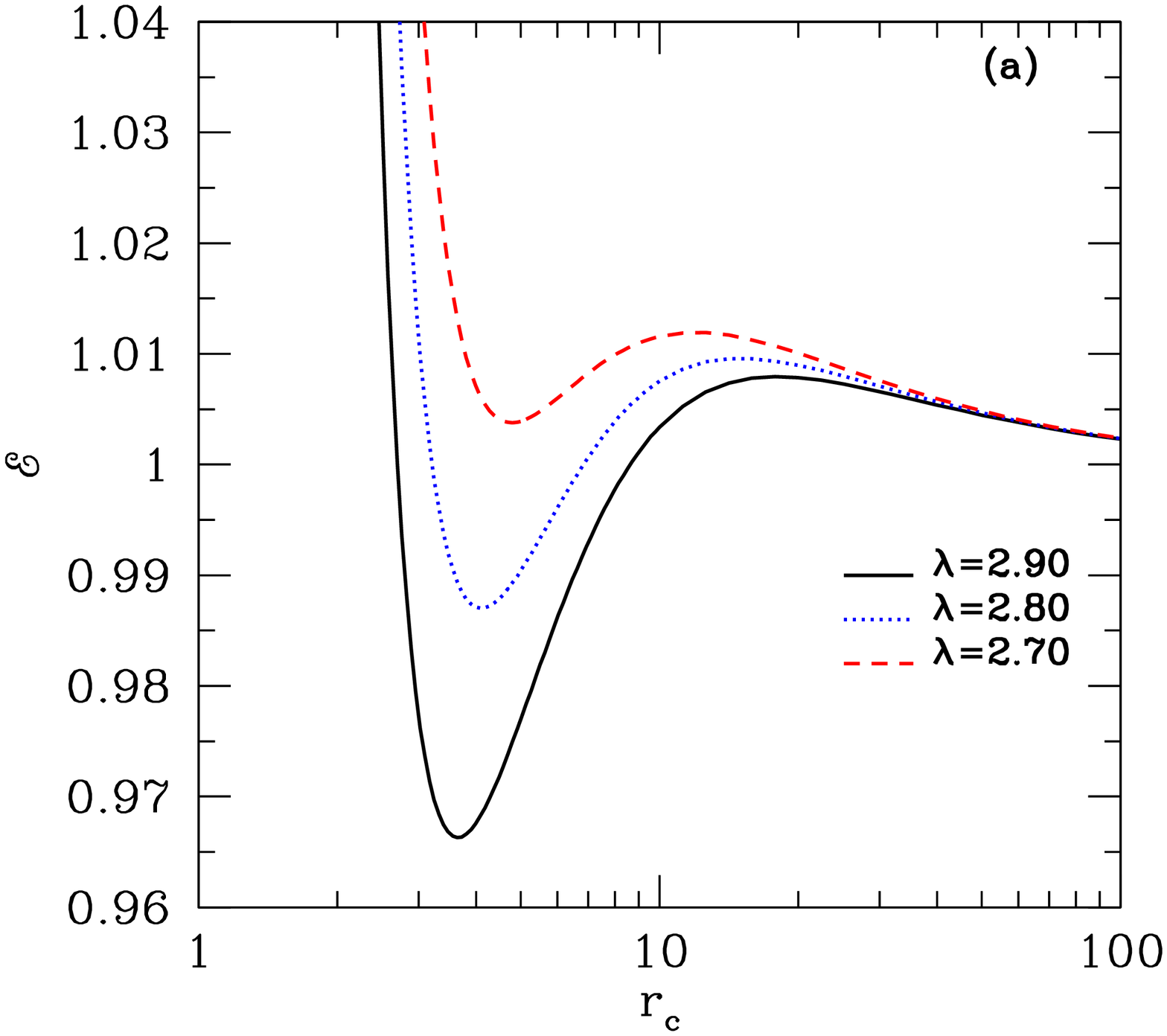}
	\includegraphics[scale=0.4]{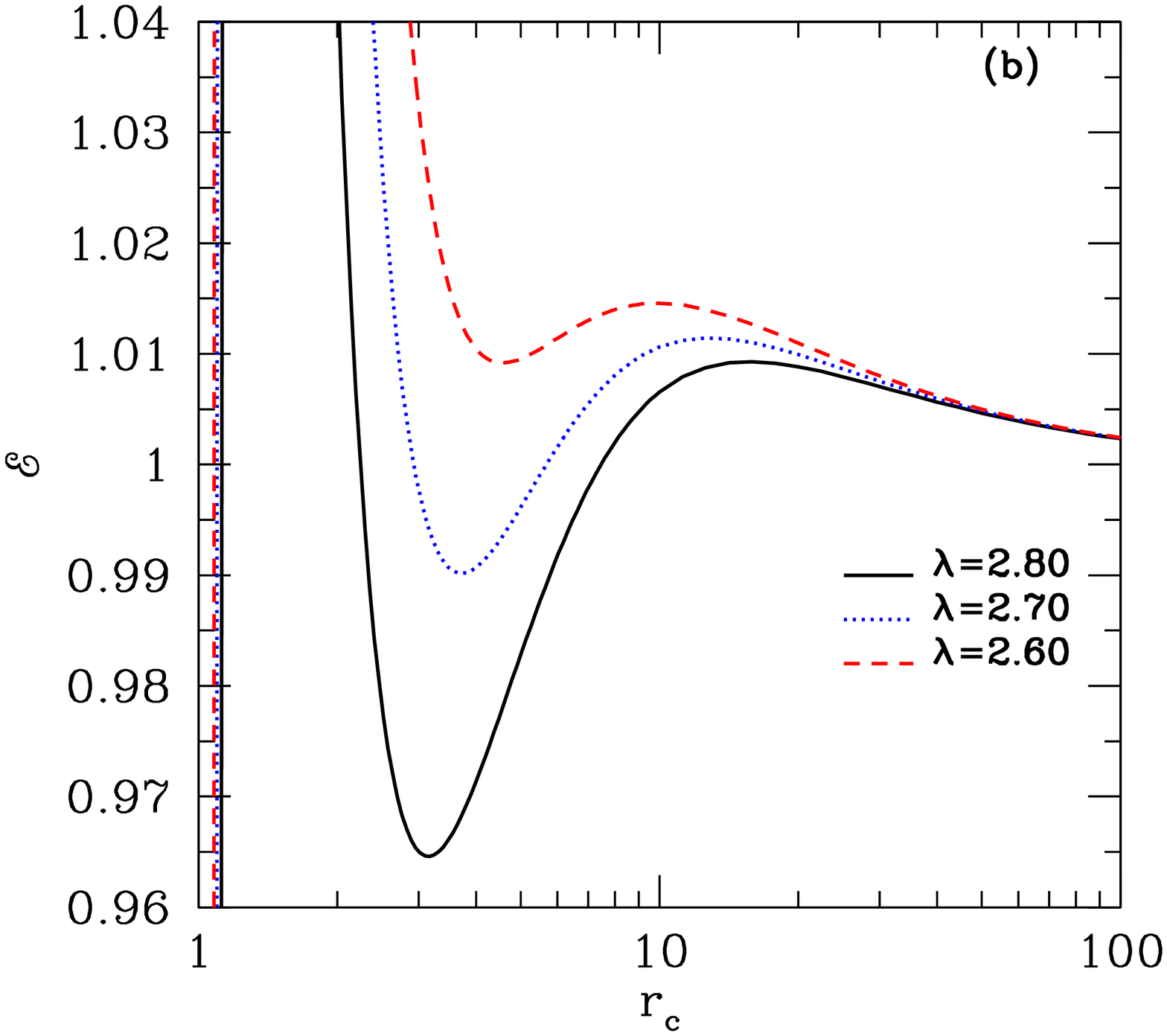}
	\includegraphics[scale=0.4]{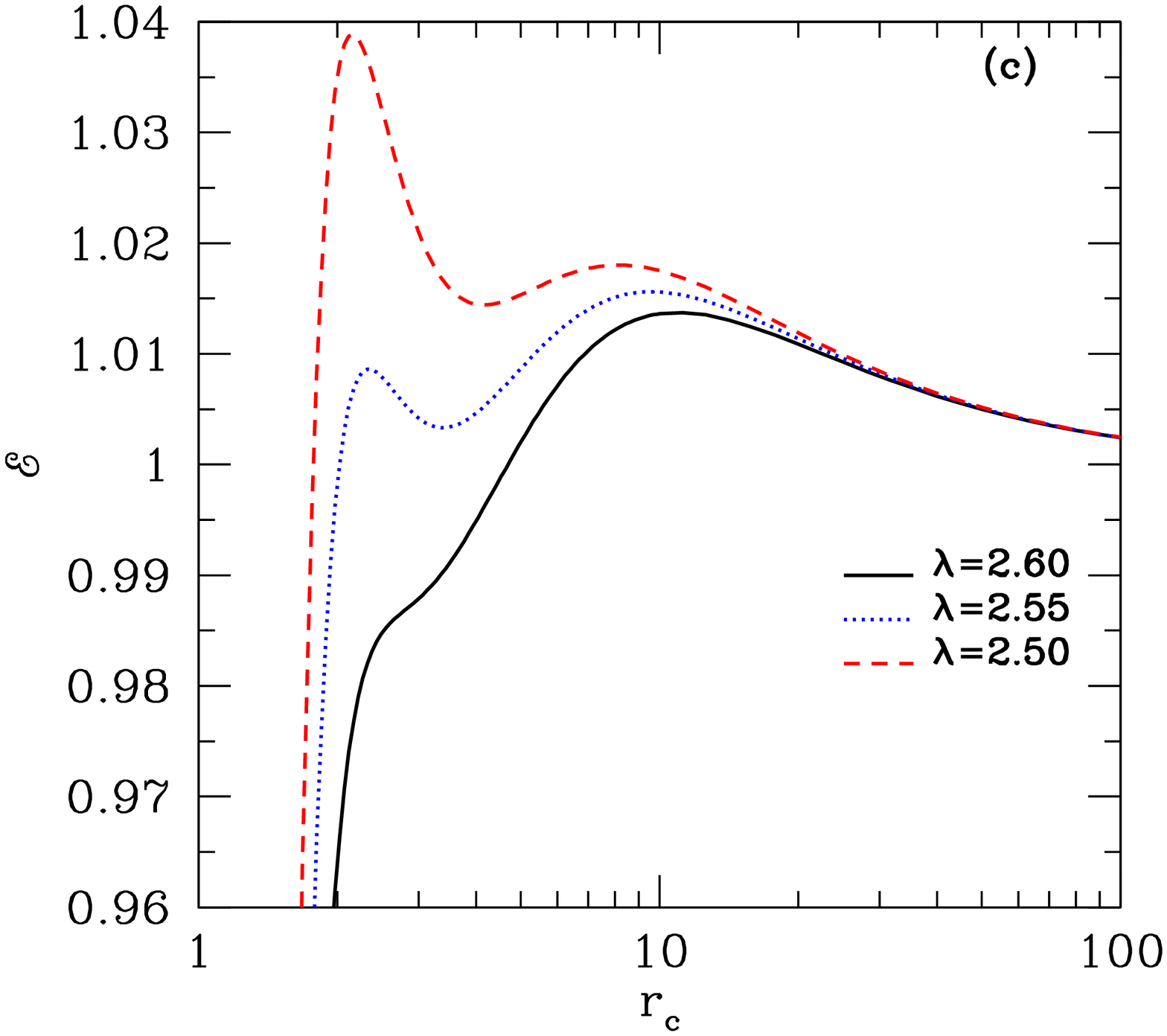}
	\includegraphics[scale=0.4]{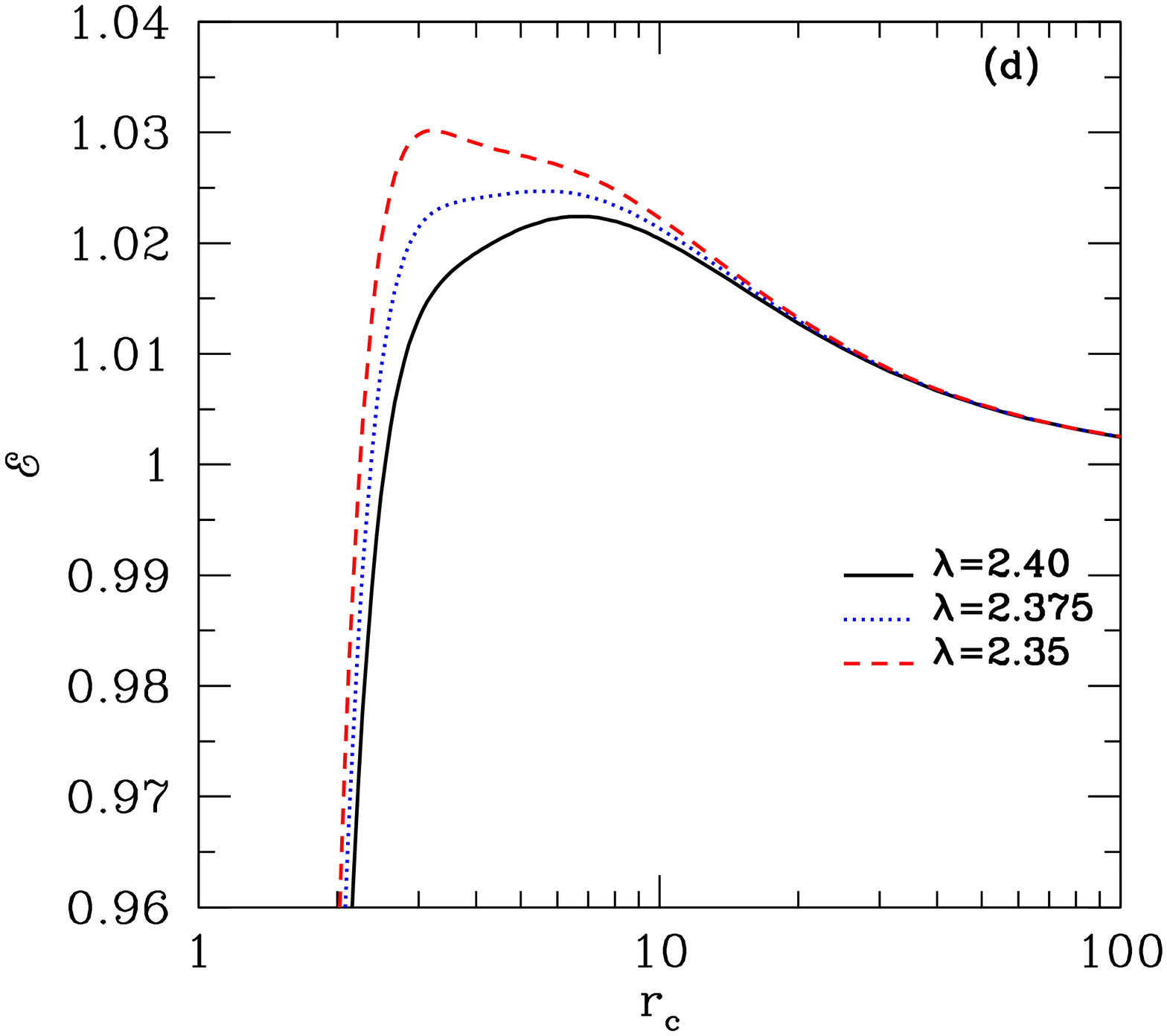}
	\caption{Plot of energy $({\cal E})$ as a function of critical point location $(r_c)$. Here we fix $n=1.24$. In each panel (a-d), solid (black), dotted (blue) and dashed (red) curves denote results for different $\lambda$ values which are marked.  We choose $a_{\rm k}=1.55$ for panel (a), $a_{\rm k}=1.60$ for panel (b), $a_{\rm k}=1.65$ for panel (c), and  $a_{\rm k}=1.70$ for panel (d), respectively. See text for details. 
		}
	\label{fig:rc_E_L}
\end{figure*}

In Fig. \ref{fig:rc_E_L}, we present the variation of flow energy (${\cal E}$) as a function of critical point location ($r_{\rm c}$) for different angular momentum ($\lambda$). In the figure, we choose NUT and Kerr parameters as $(n,a_{\rm k})=(1.24, 1.55)$ in panel (a), ($1.24, 1.60$) in panel (b), ($1.24, 1.65$) in panel (c), and ($1.24, 1.70$) in panel (d).  In panel (a), the energy variation plotted using solid (black), dotted (blue) and dashed (red) are obtained for angular momenta $\lambda=2.90$, $2.80$ and $2.70$, respectively and they are marked. Here, we consider $n, a_{\rm k}$ in such a way  that it yields the KTN black hole spacetime. This results apparently help us to understand how the properties of the critical point alter as the spacetime geometry is changed from KTN black hole to KTN naked singularity. From panel (a) it is clear that for a given set of $\lambda$ and ${\cal E}$, the flow may possess maximum of three critical points and minimum of one critical point. When multiple critical points are present, one of them is necessarily O-type in nature \cite{Fukue1987,Chakrabarti1989}.
In panel (b), we keep the NUT parameter same as in panel (a) ($i.e.$, $n=1.24$) and increase the Kerr parameter to $a_{\rm k}=1.60$, so that the spacetime contains naked singularity.
 Here, the results plotted using solid (black), dotted (blue) and dashed (red) curves are obtained for $\lambda=2.80$, $2.70$, and $2.60$, respectively. We find that near the origin, a new critical point is appeared which was absent for black hole spacetime. In reality, this critical point is invisible for black hole as it always remains inside the horizon \cite{Das-etal2001}. Hence, for a given set of $\lambda$ and ${\cal E}$, flow can have maximum of four critical points for KTN naked singularity. Among them, the innermost critical point is always O-type, and the flow can have a maximum of two saddle type critical points. More precisely, the flow contains critical points in a systematic order as O-type --- saddle type --- O-type --- saddle type with the shift of the location of the critical point away from the origin.
In panel (c), we choose $n=1.24$ and $a_{\rm k}=1.65$ where spacetime represents KTN naked singularity and the amount of spacetime deformation is more compared to (b). In the plot, solid (black), dotted (blue) and dashed (red) curves are for $\lambda=2.60$, $2.55$, and $2.50$, respectively. We find that there exists the ranges of $\lambda$ and ${\cal E}$ for multiple critical points which is reduced compared to the results presented in panel (b). Moreover, we observe that the locations of the critical points are in general shifted outwards from the origin.
Finally in panel (d), we fix $n=1.24$ and $a_{\rm k}=1.70$. This also represents the KTN naked singularity having stronger deformation of spacetime. Here, solid (black), dotted (blue) and dashed (red) curves are obtained for $\lambda=2.40$, $2.375$, and $2.35$, respectively. We find that in this limit, flow possesses at most two critical points where the inner one is O-type and the other is saddle type. We point out that possibly for the first time to the best of our knowledge, this observation is explored in the present work which is not seen for black hole spacetime.

\subsection{Parameter space for Multiple saddle type critical points}

\begin{figure}
	\includegraphics[scale=0.4]{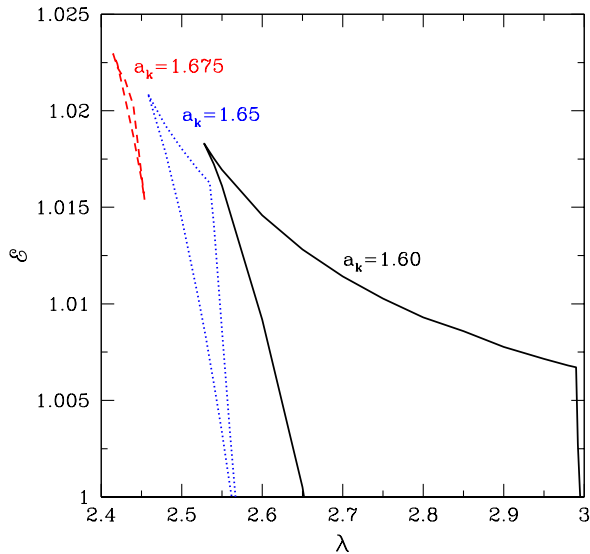}
	\caption{Modification of the parameter space for multiple critical points as $a_{\rm k}$ is increased. Here, we choose $n=1.24$. Solid (black), dotted (blue) and dashed (red) boundaries are obtained for $a_{\rm k}=1.60, 1.65$, and $1.675$, respectively. See text for details.
	}
	\label{fig:L_E_ak_NS}
\end{figure}

From the discussion presented in \S VA, it is clear that flow may harbor maximum of four critical points depending on the $a_{\rm k}$ and $n$ values. To quantify this, we identify the region of the parameter space in the $\lambda-{\cal E}$ plane for a given set of ($a_{\rm k}, n$) that allows the flow to possess at least two saddle type critical points. For that we fix $n=1.24$ and calculate the parameter space for multiple saddle type critical points for various $a_{\rm k}$ values. We present the results in Fig. \ref{fig:L_E_ak_NS}, where the region identified with solid (black), dotted (blue) and dashed (red) boundaries are obtained for $a_{\rm k}=1.60, 1.65$, and $1.675$, respectively. We observe that parameter space shrinks as well as shifts towards the lower $\lambda$ and higher ${\cal E}$ side as the $a_{\rm k}$ is increased. This eventually indicates that for a given $n$, the possibility of having multiple saddle type critical points in a flow is reduced when $a_{\rm k}$ is increased.

\subsection{Flow solutions of different kind}

In order to obtain flow solution around a naked singularity, we first choose $a_{\rm k}$ and $n$ values such that $1-a_{\rm k}^2>n^2$. Then, we calculate the critical points corresponding to flow energy ${\cal E}$ and angular momentum $\lambda$. Following the criteria to classify the nature of the critical points, we identify the saddle type critical points and calculate the flow solutions passing through it. In the next subsections, for the purpose of representation, we choose $a_{\rm k}=1.60$ and $n=1.24$, and obtain different types (altogether five types) of flow solutions for various sets of ${\cal E}$ and $\lambda$ values.

\vskip 0.3cm
\noindent {\bf A-type solutions:}

\begin{figure}
	\includegraphics[scale=0.4]{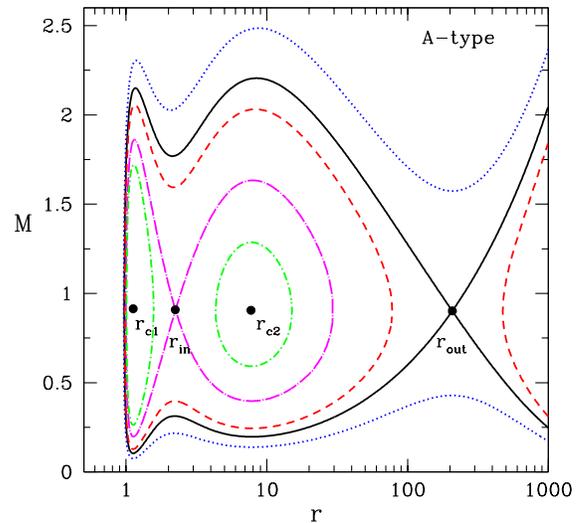}
	\caption{Plot of Mach number ($M$) as function of radial coordinate ($r$). Here, we choose $a_{\rm k}=1.60$, $n=1.24$, ${\cal E}=1.001$, and $\lambda=2.80$. Filled circles denote the critical points which are marked. The contours are of constant entropy accretion rate ($\dot {\cal M}$) which are indicated by different line styles. See text for details.
	}
	\label{fig:r_M_Atype}
\end{figure}

In Fig. \ref{fig:r_M_Atype}, we present the variation of Mach number ($M$) with radial coordinate ($r$). Here, we fix $a_{\rm k}=1.60$, $n=1.24$, ${\cal E}=1.001$ and $\lambda=2.80$, and obtain four critical points.
Among them two are saddle type critical points located at $r_{\rm in}=2.2362$ and $r_{\rm out}=207.7773$, and the other two are O-type critical points located at $r_{\rm c1}=1.1229$, $r_{\rm c2}=7.7180$, respectively. Note that the nature of the critical points appears as O-type --- saddle type --- O-type ---  saddle type in ascending order. In order to obtain the flow solution passing through $r_{\rm out}$, we calculate the radial velocity gradient $(dv/dr)_{\rm c}$ at $r_{\rm out}$ which yields two real values; one is positive and other is negative. Using the negative values of $(dv/dr)_{\rm c}$, first we integrate equation (18) and (21) inward towards the naked singular point and then outward up to the outer edge of the disc ($r_{\rm edge}$, usually the large distance). Finally, we join this two parts to obtain a complete branch of solution. Here, we choose $r_{\rm edge}=1000$. Considering the positive values of $(dv/dr)_{\rm c}$, we repeat the above procedure to obtain the other branch of the solution. In the figure, these two branches of the solution is plotted using solid (black) curve. It is noteworthy that the entropy accretion rate (${\dot {\cal M}}$) of a given solution always remains constant. We calculate the entropy accretion rate of the above flow solutions and obtain as ${\dot {\cal M}_{\rm out}} = 2.7599 \times 10^7$. Next, we calculate the flow solutions passing through the inner critical point ($r_{\rm in}$) in the same way as in the case of solutions passing through the outer critical point ($r_{\rm out}$). The noticeable difference here is that flow does not extend up to the outer edge of the disc, instead it becomes closed in between $r_{\rm c1}$ and $r_{\rm out}$. These solutions are plotted using dot-big-dashed (magenta). For this solution we find $\dot{\cal M}_{\rm in}=5.1207 \times 10^7$. Now, keeping all the remaining flow parameters unchanged, if we consider $\dot{\cal M}$ other than $\dot{\cal M}_{\rm in}$ or $\dot{\cal M}_{\rm out}$, flow solution does not possess any critical point. In that case, one can start integration of the equations (18) and (21) from any radial coordinate of interest. For example, when $\dot{\cal M} =1.9777\times 10^7$, we calculate the radial velocity ($v$) and flow temperature ($\Theta$) at $r_{\rm edge}=1000$ using equation (17) and employing them, we obtain the solution depicted by doted (blue) curve. Similarly, solutions plotted using dashed (red) and dot-small-dashed (green) curves are obtained for  $\dot{\cal M} = 3.3724 \times 10^7$ and $\dot{\cal M} = 6.7397 \times 10^7$, respectively.
We also observe that there exists a region around the naked singularity which remains inaccessible to the flow. We conjecture that during accretion, flow is expected to pile up there and tended to rotate along a surface around the naked singularity which we call as the {\it naked surface}. Since the acceptable flow solution connects the central object and the outer edge of the disc (in case of black hole, it is event horizon to the outer edge of the disc), solutions plotted with solid (black) and dotted (blue) curves are physically acceptable. However, since the entropy content of the dotted (blue) solution is lower than the solid (black) one, nature therefore favors the flow solutions passing through $r_{\rm out}$ only. 

\vskip 0.3cm
\noindent {\bf W-type solutions:}

\begin{figure}
	\includegraphics[scale=0.4]{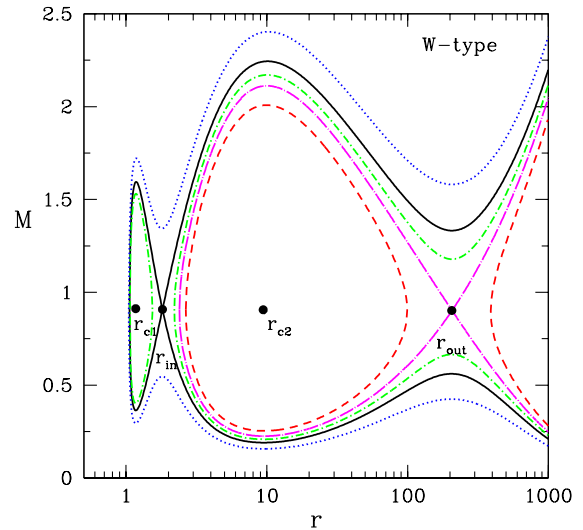}
	\caption{Plot of Mach number ($M$) as function of radial coordinate ($r$). Here, we choose $a_{\rm k}=1.60$, $n=1.24$, ${\cal E}=1.001$, and $\lambda=2.90$. Filled circles denote the critical points which are marked. The contours are of constant entropy accretion rate ($\dot {\cal M}$) which are indicated by different line styles. See text for details.
		}
	\label{fig:r_M_Wtype}
\end{figure}

Here, we choose the input parameters as $a_{\rm k}=1.60$, $n=1.24$, ${\cal E}=1.001$ and $\lambda=2.90$, and obtain the critical points as $r_{\rm c1}=1.1702$ (O-type), $r_{\rm in}=1.8138$ (saddle type), $r_{\rm c2}=9.4241$ (O-type), and $r_{\rm out}=206.0436$, respectively. In order to calculate the flow solutions, we follow the same procedure as in Fig. \ref{fig:r_M_Atype} and depict the obtained results ($M$ vs. $r$) in Fig. \ref{fig:r_M_Wtype}. The entropy accretion rate (${\dot {\cal M}}$) corresponding to the flow solutions drawn using dotted (blue), solid (black), dot-small-dashed (green), dot-big-dashed (magenta) and dashed (red) curves are calculated as $1.9594 \times 10^7$, $2.3597 \times 10^7$, $2.5688 \times 10^7$, $2.7503 \times 10^7$, and $3.0846 \times 10^7$, respectively. Note that the flow solution passing through $r_{\rm out}$ (dot-big-dashed, magenta) fails to join with the {\it naked surface}, however, solution passing through $r_{\rm in}$ (solid, black) smoothly connects the {\it naked surface} with the outer edge of the disc. Moreover, this solution is preferred over other solutions as it has high entropy content. 

\vskip 0.3cm
\noindent {\bf I-type solutions:}

\begin{figure}
	\includegraphics[scale=0.4]{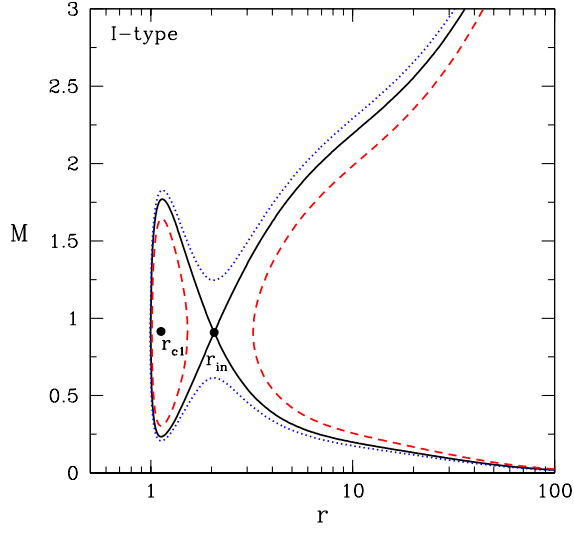}
	\caption{Plot of Mach number ($M$) as function of radial coordinate ($r$). Here, we choose $a_{\rm k}=1.60$, $n=1.24$, ${\cal E}=1.03$, and $\lambda=2.80$. Filled circles denote the critical points which are marked. The contours are of constant entropy accretion rate ($\dot {\cal M}$) which are indicated by different line styles. See text for details.
		}
	\label{fig:r_M_Itype}
\end{figure}

We continue our study of finding flow solutions and choose the input parameters as $a_{\rm k}= 1.60$, $n=1.24$, ${\cal E}=1.030$ and $\lambda=2.80$. Here, we find that only two critical points exist: one of them is O-type ($r_{\rm c1}$) and the other is saddle type ($r_{\rm in}$). We calculate the flow solutions following the procedure as in Fig. \ref{fig:r_M_Atype}. We observe that flow solution with ${\dot {\cal M}}= 6.9873 \times 10^7$ passes through $r_{\rm in}=2.0610$ and connects the {\it naked surface} to the outer edge of the disc which is shown using solid (black) curve in Fig. \ref{fig:r_M_Itype}. Other solutions obtained for ${\dot {\cal M}}=6.2090 \times 10^7$ and $8.8773 \times 10^7$ are plotted with dotted (blue) and dashed (red) curves as shown in the figure. As before solution passing through $r_{\rm in}$ is preferred as it has higher ${\dot {\cal M}}$. 

\vskip 0.3cm
\noindent {\bf O-type solutions:}

\begin{figure}
	\includegraphics[scale=0.4]{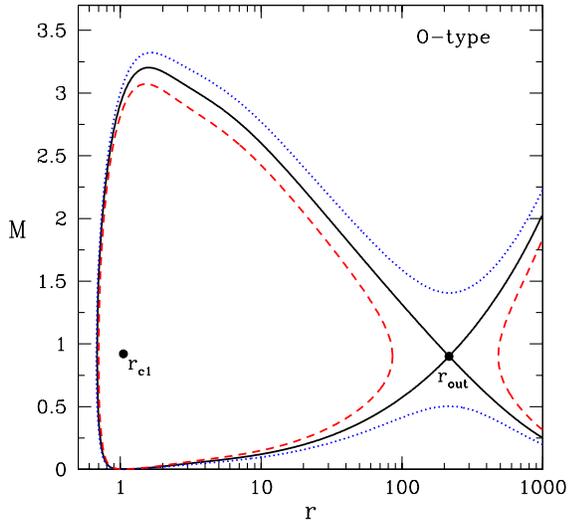}
	\caption{Plot of Mach number ($M$) as function of radial coordinate ($r$). Here, we choose $a_{\rm k}=1.60$, $n=1.24$, ${\cal E}=1.001$, and $\lambda=2.22$. Filled circles denote the critical points which are marked. The contours are of constant entropy accretion rate ($\dot {\cal M}$) which are indicated by different line styles. See text for details.
	}
	\label{fig:r_M_Otype}
\end{figure}

In this case, we choose the input parameters as $a_{\rm k}=1.60$, $n=1.24$, ${\cal E}=1.001$ and $\lambda=2.22$ and find two critical points. Here, we find that the flow solutions are very much similar in character as in Fig. \ref{fig:r_M_Itype}, except the saddle-type critical point ($r_{\rm out}$) forms far away from the {\it naked surface} and flow solution passing through $r_{\rm out}=216.2226$ is extended from the {\it naked surface} to the outer edge. In Fig. \ref{fig:r_M_Otype}, the solutions plotted using dotted (blue), solid (black) and dashed (red) curves are obtained for ${\dot {\cal M}}=2.2567 \times 10^7$, $2.8095 \times 10^7$, and $3.4239 \times 10^7$, respectively.

\vskip 0.3cm
\noindent {\bf I$^{*}$-type solutions:}

\begin{figure}
	\includegraphics[scale=0.4]{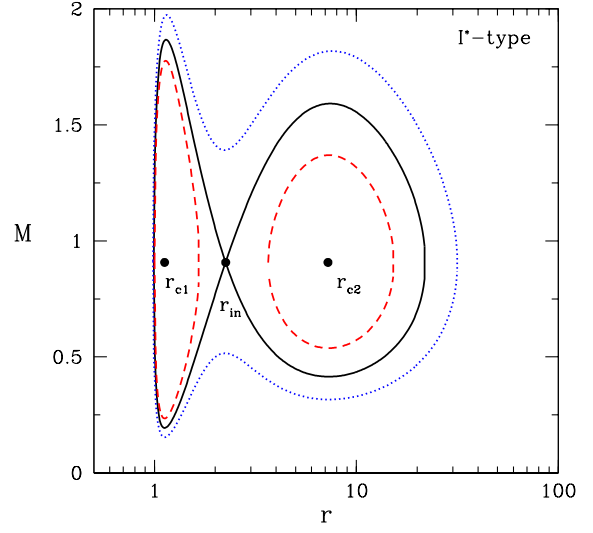}
	\caption{Plot of Mach number ($M$) as function of radial coordinate ($r$). Here, the solutions are obtained for bounded energy (${\cal E} < 1$) where flow parameters are chosen as $a_{\rm k}=1.60$, $n=1.24$, ${\cal E}=0.999$, and $\lambda=2.80$, respectively. Filled circles denote the critical points which are marked. The contours are of constant entropy accretion rate ($\dot {\cal M}$) which are indicated by different line styles. See text for details.
		}
	\label{fig:r_M_I*type}
\end{figure}

Here, we present the flow solutions for bounded energies, $i. e., ~{\cal E}<1$. For that, we choose the input parameters as $a_{\rm k}= 1.60$, $n=1.24$, ${\cal E}=0.999$ and $\lambda=2.80$ and obtain three critical points. Among them, two are O-type ($r_{\rm c1}$ and $r_{\rm c2}$) and the remaining one is saddle type ($r_{\rm in}$). As before, we calculate the flow solutions following the procedure mentioned above while generating Fig. \ref{fig:r_M_Atype} and depict all the solutions in Fig. \ref{fig:r_M_I*type}. For ${\dot {\cal M}}= 5.00129\times10^7$, flow solution passes through $r_{\rm in}=2.2526$ and becomes closed which is shown using solid (black) curve. Solution of this kind is physically delusive as it fails to produce global accretion solution connecting the {\it naked surface} to the outer edge of the disc. Other solutions which are not transonic in nature, are obtained for ${\dot {\cal M}}=4.00129 \times 10^7$ and $6.00129 \times 10^7$ and we plot them using dotted (blue) and dashed (red) curve as shown in the figure.

\section{Discussion and Conclusions}

In this work, we study the properties of the accretion flow in a general axisymmetric KTN spacetime. This spacetime either describe black hole or naked singularity depending on the choice of Kerr parameter $(a_{\rm k})$ and NUT parameter $(n)$. We consider the relativistic hydrodynamic equations that govern the flow motion and solve them to obtain the flow solutions around the black holes or naked singularities in the steady state limit. We examine the role of $a_{\rm k}$ and $n$ in deciding the nature of the critical points as well as the flow solutions. We present our finding point wise below.

\begin{enumerate}
	\item For KTN black hole with fixed $a_{\rm k}$, there exists a range of $n$ that admits maximum of three critical points. Among them, the critical point that forms close to the horizon is always saddle type. Beyond this range, flow is left with only one critical point (see Fig. \ref{fig:rc_E_n}). 
	
	\item We calculate all possible transonic flow solutions around a KTN black hole and separate the  parameter space in $\lambda-{\cal E}$ plane according to the nature of the flow solutions (see Fig. \ref{fig:L_E_r_M}). We also observe that the nature of the flow solutions changes as $n$ is varied (see Fig. \ref{fig:r_M_n}). Considering this, we study the modification of $\lambda - {\cal E}$ parameter space for multiple critical points and find that for a given $a_{\rm k}$, as $n$ is increased, the parameter space is shifted towards the higher angular momentum and lower energy domain (see Fig. \ref{fig:L_E_n}). On the other hand, when $a_{\rm k}$ is increased keeping $n$ fixed, the shift of the parameter space happens in the lower angular momentum and higher energy sides (see Fig. \ref{fig:L_E_ak}). These findings suggest that $a_{\rm k}$ and $n$ respond in opposite way in determining the parameter space for multiple critical points. Overall, it appears that the NUT parameter ($n$) effectively shields the black hole rotation for flows accreting on to them.
	
	\item It may be noted that, for KTN spacetime, $a_{\rm k}>1$ are possible as opposed to the usual Kerr black holes where $a_{\rm k} < 1$. We therefore study the $\lambda-{\cal E}$ parameter space for multiple critical points considering KTN black hole having Kerr parameter $a_{\rm k} > 1$. Considering the various combination of $a_{\rm k}$ and $n$ values, we obtain a fixed event horizon $r_{\rm H}$ (see equation (2)), and obtain the multiple critical point parameter space. We observe that the parameter space is very much dependent on $r_{\rm H}$ when $a_{\rm k}$ and $n$ values are small with respect to unity, however, it tends to become independent on $r_{\rm H}$ when both $a_{\rm k}$ and $n$ is very large (see Fig. \ref{fig:L_E_ak_n}).
	
	\item We compute the maximum luminosity ($L_{\rm max}$) to be emitted by the accretion flow considering the Bremsstrahlung radiative process active in the flow. We find that $L_{\rm max}$ in general decreases with the increase of $n$ irrespective to the $a_{\rm k}$ values (see Fig. \ref{fig:n_Lum_ak}).

	\item We examine the critical point properties considering the naked singularity and reveal that flow may possess maximum of four critical points. When flow contains four critical points, two of them must be saddle type critical points (see Fig. \ref{fig:rc_E_L}). We calculate $\lambda-{\cal E}$ parameter space for multiple saddle type critical points and find that the parameter space shrinks and shifted towards lower $\lambda$ and higher ${\cal E}$ side as $a_{\rm k}$ is increased (see Fig. \ref{fig:L_E_ak_NS}). We further obtain the all possible transonic flow solutions where we find that flow tends to reach an imaginary surface called as {\it naked surface} avoiding the origin of the naked singularity (Figs. \ref{fig:r_M_Atype}-\ref{fig:r_M_I*type}).

\end{enumerate}

Finally, we argue that our formalism may be used to predict the possible range of NUT parameter ($n$) in the astrophysical context. In order to do that, one requires the knowledge of the source luminosity, source mass and source spin, respectively (see Fig.  \ref{fig:n_Lum_ak}). To keep our discussion simple, in this work, we only considered Bremsstrahlung emission process neglecting the other radiative processes, namely synchrotron emission, Compton emission, etc., although they are expected to play a role in determining the accretion disc luminosity. Therefore, in order to constrain the range of $n$, a rigorous study is indispensable, involving all the emission processes, which we intend to consider as a future work and plan to report elsewhere.

\section*{Acknowledgments}
 
 All authors thank Indian Institute of Technology Guwahati, India for providing infrastructural support to carry out this work. ID thanks the Max Planck Partner group grant (MPG-01) for financial support.

%\bibliography{reference}
\label{lastpage}
\end{document}